\documentclass[epj]{svjour}
\usepackage{graphics}
\usepackage{epstopdf}
\usepackage{amsmath}
\usepackage{amssymb}
\usepackage{color}
\begin{document}
\title{Boundary flow of viscoelastic polyelectrolyte solutions}
\author{Chlo\'e Barraud\inst{1} \and Benjamin Cross\inst{1}  \and Cyril Picard\inst{1}
 \and Fr\'ed\'eric Restagno\inst{2} \and Liliane L\'eger\inst{2} \and Elisabeth Charlaix\inst{1} 
\thanks{\emph{Present address:} elisabeth.charlaix@univ-grenoble-alpes.fr}%
}                     
\offprints{}          
\institute{Univ. Grenoble Alpes, CNRS, LIPhy, F-38000 Grenoble, France \and Laboratoire de Physique des Solides, CNRS, Univ. Paris-Sud, Universit\'e Paris-Saclay, 91405 Orsay Cedex, France}
\date{Received: date / Revised version: date}
%
\abstract{
We report an investigation of the equilibrium and dynamic properties of polyelectrolyte solutions confined between platinum surfaces with a dynamic Surface Force Apparatus. 
The polyelectrolyte adsorbs on the surfaces in a dense compact layer bearing a surface charge in good agreement with the theoretical predictions. The flow of the solution on this charged adsorbed layer is probed over four decades of spatial scales and one decade of frequency by dynamic measurements.  At distances larger than the hundredth of nanometers, the flow of the viscoelastic solution is well described by a partial slip boundary condition. We show that the wall slip is quantitatively described by an interfacial friction coefficient, according to the original Navier's formulation, and not by a slip length. At smaller distance the partial slip model overestimates the solution mobility, and we observe the presence of a low viscosity layer coating the surfaces. The viscosity and thickness of this boundary layer are directly resolved, and found to be independent on shear-rate, frequency, and confinement. We discuss the thickness of the low viscosity layer in terms of the structural length of the semi-dilute solution and the Debye length screening the adsorbed layer charge. 
%
} 
\maketitle
\section{Introduction}
\label{intro}
When a solid surface is put in contact with a polymer solution, different situations can occur depending on the polymer-surface interactions \cite{de_gennes_polymer_1981}. If the attraction is large, polymer will adsorb on the surface. On the contrary, the attraction between the polymer chains close to the surface is low enough, a depletion layer can be obtained leading to a zone of polymer concentration lower than in the bulk. Neutron reflectivity has been shown to be a unique tool to study the structure of the interfacial layer \cite{auvray_neutron_1986,auvray_polymer_1988,auvray_irreversible_1992,auvray_structure_1992}. The structural characterization of the interfacial layer has been the subject of pioneering work by L. Auvray. In particular, in the case of a semi-dilute solution of polymer chains of $N$ monomers, which $a$ is the monomer size,  at concentration $\phi$, he measured the adsorbed layers thickness $h=aN^{1/2}\phi^{7/8}$ as predicted by the theory \cite{auroy_building_1991,auvray_irreversible_1992}. The same experimental neutron reflectivity setup has been used to characterize the depletion layer. In  the  dilute regime,  the  
thickness of  the depletion  layer  is of the
order of the radius of  gyration  of the free  polymer  coil. As the concentration increases and 
approaches  coil  overlap in the semi-dilute regime, the depletion layer decreases significantly 
to the order of the bulk polymer correlation length, $\xi$ \cite{lee_direct_1991}.
 
There are many practical consequences of the existence of either an adsorbed or a depletion layer at the interface between a polymer solution. If the structure of the interfaces at equilibrium is now well established, the mechanical properties of the interfaces is still a only partially understood question. A way to probe mechanically the interfacial properties of the interface when a polymer solution is put into contact with a wall is generally to push a fluid on this solution to create a continuous flow over the interfacial layer.  The effect of the surface properties can thus be incorporated into a continuum description via a hydrodynamic  boundary  condition  (h.b.c.)  to  the  Navier-Stokes equation to describe the fluid-solid friction through a slip at the interface.   The  slip  boundary  condition  was formulated by Navier in 1823 in the case of Newtonian liquids \cite{Navier} as a balance between viscous
stress and friction stress at the surface:
\begin{equation}
\eta\frac{\partial v}{\partial z}_{|z=0}= \lambda v(0) 
\end{equation}
where $\eta$ denotes the shear viscosity of the liquid and  $\lambda$  the friction coefficient at the interface.

The amount of slip is usually quantified
by  the  slip  length $b=\eta/\lambda$.  A negative slip length ($b<0$) is associated to the presence of an adsorbed  layer which corresponds to an hydrodynamically dead layer close to the solid-liquid interface. This has been shown with different experimental techniques based on 
the reduction of flux of solvent when a polymer is adsorbed on the wall of a suitable flow 
channel.  Various materials and configurations have been used, for  example, sintered glass disks, porous membranes, 
single glass capillaries, capillary arrays, the narrow channel of a Surface Forces Apparatus \cite{cann1994behavior,cohen_stuart_hydrodynamics_1984,li2014slip}. 

On the contrary, a depletion layer is associated to a layer of higher mobility close to the surface \cite{barnes_review_1995}. This can be associated to a positive slip length that is often only considered as an apparent slip \cite{neto_boundary_2005,sanchez-reyes_interfacial_2003,cayer-barrioz_drainage_2008,cuenca_submicron_2013,fang_dna_2005} instead of a real slip. Indeed, in the
presence of a thin layer of thickness $\delta$ depleted in polymer molecules, the apparent slip length is \cite{VinogradovaLA95,LaugaBrenner}:
\begin{equation}\label{equ_depletion}
b=\delta \left ( \frac{\eta}{\eta_{\rm s}}-1 \right ),
\end{equation}
where $\eta$ and $\eta_{\rm s}$ are the solution and
solvent viscosity. The size of this layer has
been in debate since the 1980s \cite{barnes_review_1995}. It can be either close to the radius of gyration for dilute polymer solutions or the correlation length for semi-dilute solutions \cite{chauveteau_concentration_1984} or even greater than $R_{\rm g}$ or $\xi$ if a migration mechanism occurs. More recently, this as received larger attention thanks to the development of micro to nano-fluidics where people have studied the migration of DNA solutions \cite{fang_concentration_2007,fang_dna_2005,boukany_exploring_2009,boukany_molecular_2010,cuenca_submicron_2013} or developed numerical models \cite{fan2003microchannel,graham2011,millan2007pressure,kekre2010role}.

Nevertheless, it should be noticed that polymer solutions of long polymer chains can present a viscoelastic character leading, in the linear regime to a complex viscosity.The measurement of a slip length from Eq. \ref{equ_depletion} becomes delicate. 

In the present paper, we use a Dynamic Surface Force Apparatus to precisely measure the boundary condition, i.e. the slip length of a particular viscoelastic fluid made of semi-dilute polyelectrolyte solutions. We measure in the same experiment
the equilibrium interactions and the hydrodynamic properties in a confined  polyelectrolyte solution and the nanorheological behavior of the solutions. We measure the apparent slip at large scale compatible to a less viscous layer close to the interface. In a second time, we study directly the thickness of this low viscosity boundary  layer by mechanical measurements which will allow to propose different microscopic scenarios.

\section{Materials and methods}
\label{sec:MatMet}
\subsection{Polyelectrolyte solutions and surfaces}
\label{sec:MatMet1}
We study aqueous solutions of partially hydrolyzed
polyacrylamide
(HPAM). HPAM  is a water soluble polyelectrolyte well-known for its viscosifying properties at small concentration,  and widely used in the field of enhanced oil recovery or water purification \cite{Lake1989,Wever2011}. 
\begin{figure}[!h]
\setlength{\unitlength}{1cm}
\begin{picture}(7,2.5)(0,1)
\put(0,2.8){\line(1,0){0.4}}
\put(0.5,2.7){${\rm CH}_2$}
\put(1.2,2.8){\line(1,0){0.3}}
\put(1.6,2.7){${\rm HC}$}
\put(2.2,2.8){\line(1,0){1.2}}
\put(1.9,2.3){\line(0,1){0.3}}
\put(1.8,1.9){${\rm C}$}
\put(1.9,1.5){\line(0,1){0.3}}
\put(1.8,1.2){${\rm NH}_2$}
\put(2.1,1.95){\line(1,0){0.4}}
\put(2.1,2.05){\line(1,0){0.4}}
\put(2.6,1.9){${\rm O}$}
\put(3.5,2.7){${\rm CH}_2$}
\put(4.2,2.8){\line(1,0){0.3}}
\put(4.6,2.7){${\rm HC}$}
\put(5.2,2.8){\line(1,0){1.}}
\put(4.9,2.3){\line(0,1){0.3}}
\put(4.8,1.9){${\rm C}$}
\put(4.9,1.5){\line(0,1){0.3}}
\put(4.8,1.2){${\rm OH}$}
\put(5.1,1.95){\line(1,0){0.4}}
\put(5.1,2.05){\line(1,0){0.4}}
\put(5.6,1.9){${\rm O}$}
\put(3.6,1.2){${\rm O}^-\ {\rm or}$}
\put(0.25,1.1){\line(0,1){2}}
\put(0.25,1.1){\line(1,0){0.1}}
\put(0.25,3.1){\line(1,0){0.1}}
\put(2.95,1.1){\line(0,1){2}}
\put(2.95,1.1){\line(-1,0){0.1}}
\put(2.95,3.1){\line(-1,0){0.1}}
\put(3.0,1){${\rm p}$}
\put(3.35,1.1){\line(0,1){2}}
\put(3.35,1.1){\line(1,0){0.1}}
\put(3.35,3.1){\line(1,0){0.1}}
\put(5.95,1.1){\line(0,1){2}}
\put(5.95,1.1){\line(-1,0){0.1}}
\put(5.95,3.1){\line(-1,0){0.1}}
\put(6.0,1){${\rm m}$}
\end{picture}
\caption{Partially hydrolyzed polyacrylamide.}
\end{figure}
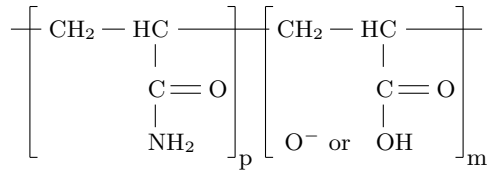

HPAM forms a linear chain of acrylamide monomers, a fraction of which presents a carboxylate group. The molar mass of the monomer is $M_A=71.25$ g/mol, and its size  is $a \approx$ 4 \r{A}. We use HPAM of molecular weight 20 10$^6$ g/mole (from SNF Flopaam 3630S)  prepared in deionized water solution, in the concentration range around 1 g/l.  In these  conditions, which were otherwise well characterized by
other groups \cite{cuenca_submicron_2013}, the carboxyl groups are fully ionized, and the fraction of charged monomer on the chains is $f =m/(m+p)\approx 0.25$. 

 In the range of concentration studied the solutions are expected to be in the  semi-dilute entangled regime. The entanglement criterion for a semi-dilute solution is $C \ge C_e = n^4 C^*$, where 
 $5\le n \le 10$ is the average number of chains with which a given chain is overlapping, and $C^*$ (in g/l) is the overlap concentration \cite{DobryninRubinstein2005}
 \begin{equation}
 C^* \approx (a fN)^{-2}l_B^{-1}M_A/\mathcal N_A10^3
 \end{equation}
Here $N=280700$ is the number of monomers in a chain, $\mathcal N_A$ is the Avogadro number, and $l_B=e^2/ 4\pi \epsilon_w \epsilon_ok_BT$ is the Bjerrum length \cite{DobryninRubinstein2005}. At the experimental temperature $T=28^o$C,   with the relative dielectric constant of water $\epsilon_w=77.22$ \cite{dielwater1956}, the Bjerrum length is  $l_B=7.2\ $\r{A} and  $C^*= 0.16 \mu$g/l   for the 20 10$^6$ g/mol HPAM.   The criterion for entangled semi-dilute solutions is fully met with a calculated number of contacts $n=50$. The correlation length in these solutions is \cite{DobryninRubinstein2005}:
\begin{equation}
\xi \approx (af)^{-1/3}l_B^{-1/6}(C \mathcal N_A 10^3/M_A)^{-1/2}
\end{equation}
 giving $\xi = 23 \ C^{-1/2}$  in nanometers,  with C in g/l.


HPAM solutions are confined between a sphere and a plane. The surfaces are made of highly smooth floated borosilicate glass with a typical roughness of 2 \r{A} over a 100 $\mu$m$^2$ area, measured by atomic force microscopy. They are then coated by a 10 nm thickness layer of chromium, and then by a 90 nm thickness layer of platinum. These metallic layers are deposited by magnetron sputtering. The typical roughness of the coating is less than 5 \r{A} over a 100 $\mu$m$^2$ area.

\subsection{Dynamic Surface Force Apparatus (dSFA)}
\label{sec:MatMet2}
The experimental setup is a new Dynamic Surface Force Apparatus (DSFA) that has recently been developed. Detailed 
description can be found in previous articles \cite{rsi2002,garcia_micro-nano-rheometer_2016}. This DSFA measures separately the relative displacement $h$ of the surfaces and the interaction force between them. The surfaces used are a plane and a sphere of radius $R=3.3$ mm. The apparatus measures the static as well as the dynamic component of the 
force when the sphere is moved towards or backwards the plane, and can therefore be used as a nanorheometer. The 
sphere can be moved in a direction normal to the plane over a distance of about 15 $\mu$m with a 
piezoelectric actuator. The same actuator allows one to superimpose sinusoidal oscillation at angular frequency $\omega$, leading to a distance between the plane and the apex of the sphere equal to: $D(t)=D+h_0\cos(\omega t)$. The average distance $D$ is varied quasi-statically during the experiments at a velocity lower than 1 nm/s). The plane is attached to a flexure hinge in order to perform a force measurement with a purely translational deflexion. The stiffness of the flexure hinge is $K=5707$~N/m, leading to a resonance frequency $f_0=118$~Hz and a quality factor $Q=500$. The frequency response of the hinge is further used to transform the displacement of the plane into the quasistatic and harmonic components of the interaction force.   The  frequency can be varied between 10 to 300~Hz. The plane displacement is measured with a Nomarski interferometer,  leading to   a  static  resolution  of 0.1~$\mu$N. The relative displacement between the sphere and the plane is directly measured with another Nomarski interferometer with a 0.1~nm resolution for the quasistatic part.  For dynamic 
measurements, the output signals of the displacement and force interferometers are connected to two digital two-phase lock-in 
amplifiers (Standford Research System SR830 DSP Lock-In Amplifier) whose reference is used to drive 
the piezoelectric actuator. In the dynamic regime, the displacement sensibility is 1~pm and the force sensibility is 10~nN. Our measurements are performed 
with a bandwidth of 1~Hz. The force exerted on the plane is written as follow: $F(t)=F_{\rm stat}(D)+F_{\rm dyn}(D,\omega,h_0)\cos(\omega t + \varphi)$. The zero-frequency component $F_{\rm stat}(D)$ is the quasi-static force at the distance $D$. It is related to the  free energy $W(D)$ of  the  polymer solution confined between two parallel plates by the Deryaguin approximation \cite{Derjaguin}:
\begin{equation}
\frac{F_{\rm stat}(D)}{R} = 2\pi W(D)
\label{deryaguin}
\end{equation}
\begin{figure}[t]
\center\resizebox{0.25\textwidth}{!}{%
  \includegraphics{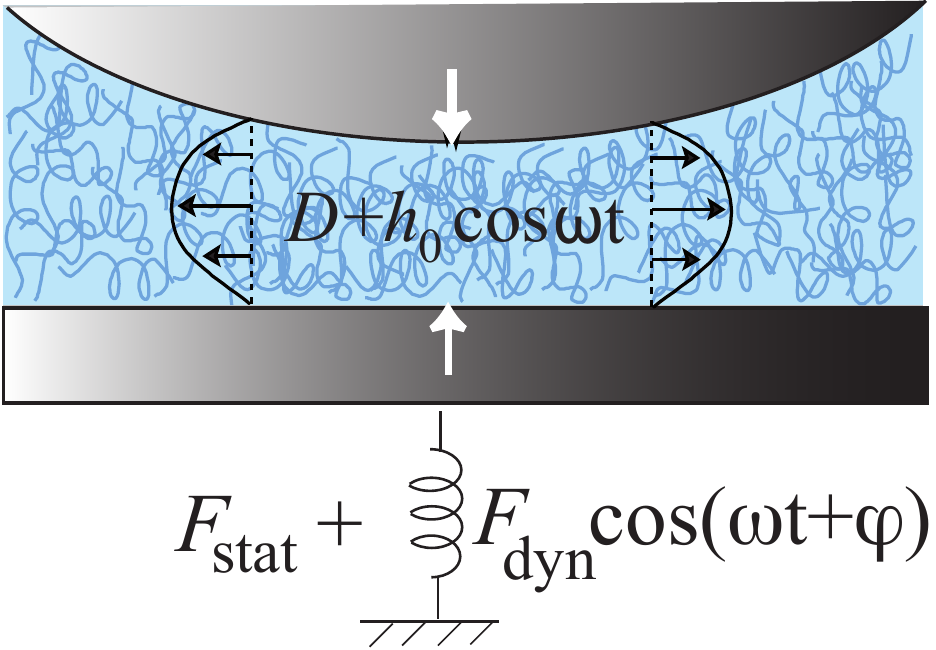}}
\caption{Schematic of the dynamic Surface Force Apparatus}
\label{fig:SFA}       
\end{figure}

The $\omega$-harmonic force component  $F_{\rm dyn}\cos(\omega t + \varphi)$ is obtained for $h_0\ll D$. In the limit of the linear response $F_{\rm dyn}$ is proportional to $h_0$ and we define 
 the  complex mobility $\tilde \mu(D,\omega)$:
\begin{equation}
\tilde \mu(D,\omega)=\frac{\dot D}{F_{\rm dyn}\exp(\text{i}\varphi)}=\frac{{\rm i} \omega h_0}{F_{\rm dyn}\exp(\text{i}\varphi)}
\end{equation}
The real and imaginary components $\mu_R$ and $\mu_I$ of the mobility measure respectively the dissipative component and the elastic component of the sphere motion under the applied dynamic force $F_{dyn}$.
The sphere mobility thus allows one to probe  the bulk linear rheology of the liquid as well as its boundary condition at the solid interface, in the linear response limit. For a visco-elastic liquid of complex viscosity $\tilde \eta = \tilde G/{\rm i }\omega$ undergoing a Navier's partial slip boundary condition of symmetric nature on the solid surfaces: 
\begin{equation}
\tilde \lambda \tilde v_{\rm slip} = \tilde \eta \frac{\partial \tilde v_x}{\partial z}
\label{Navier}
\end{equation}
the reduced mobility $R^2 \tilde \mu(D)$ at large distance $D\gg \vert \tilde \eta \vert / \vert \tilde \lambda \vert$ is asymptotically equal to \cite{cross2018}: 
 \begin{equation}
R^2\tilde \mu (D) \simeq \frac{D}{6\pi \tilde \eta}+ \frac{1}{3\pi \tilde \lambda}
\label{far_field}
\end{equation}
Thus the bulk modulus of the visco-elastic liquid can be obtained from the slope of the linear growth of the reduced mobility at large distance, whereas the boundary friction coefficient is the ordinate at origin of the linear extrapolation of this large distance behavior. The complex friction coefficient: 
\begin{equation}
\frac{1}{\tilde \lambda}=\frac{1}{\lambda_R}+\frac{{\rm i}\omega}{k}
\label{complexfriction}
\end{equation}
takes into account apparent slippage with dissipative interfacial friction $\lambda_R$, and finite interfacial compliance with a stiffness coefficient $k$. At smaller distance, the full expression of the reduced mobility with a partial slip boundary condition, and neglecting surface deformation effects, is given by the Hocking expression \cite{Hocking1973,cross2018}:
\begin{eqnarray}
R^2\tilde \mu (D) = \frac{D}{6\pi \tilde \eta f^*(\tilde \eta / \tilde \lambda D)} \cr
f^*(\tilde y) = \frac{1}{3\tilde y} \left ((1+\frac{1}{6\tilde y})\ln (1+6\tilde y)-1 \right )
\label{Hocking}
\end{eqnarray}

\section{Equilibrium properties of confined polyelectrolyte solutions}
\label{sec:static}

\begin{figure}[hb]
\center\resizebox{0.45\textwidth}{!}{%
  \includegraphics{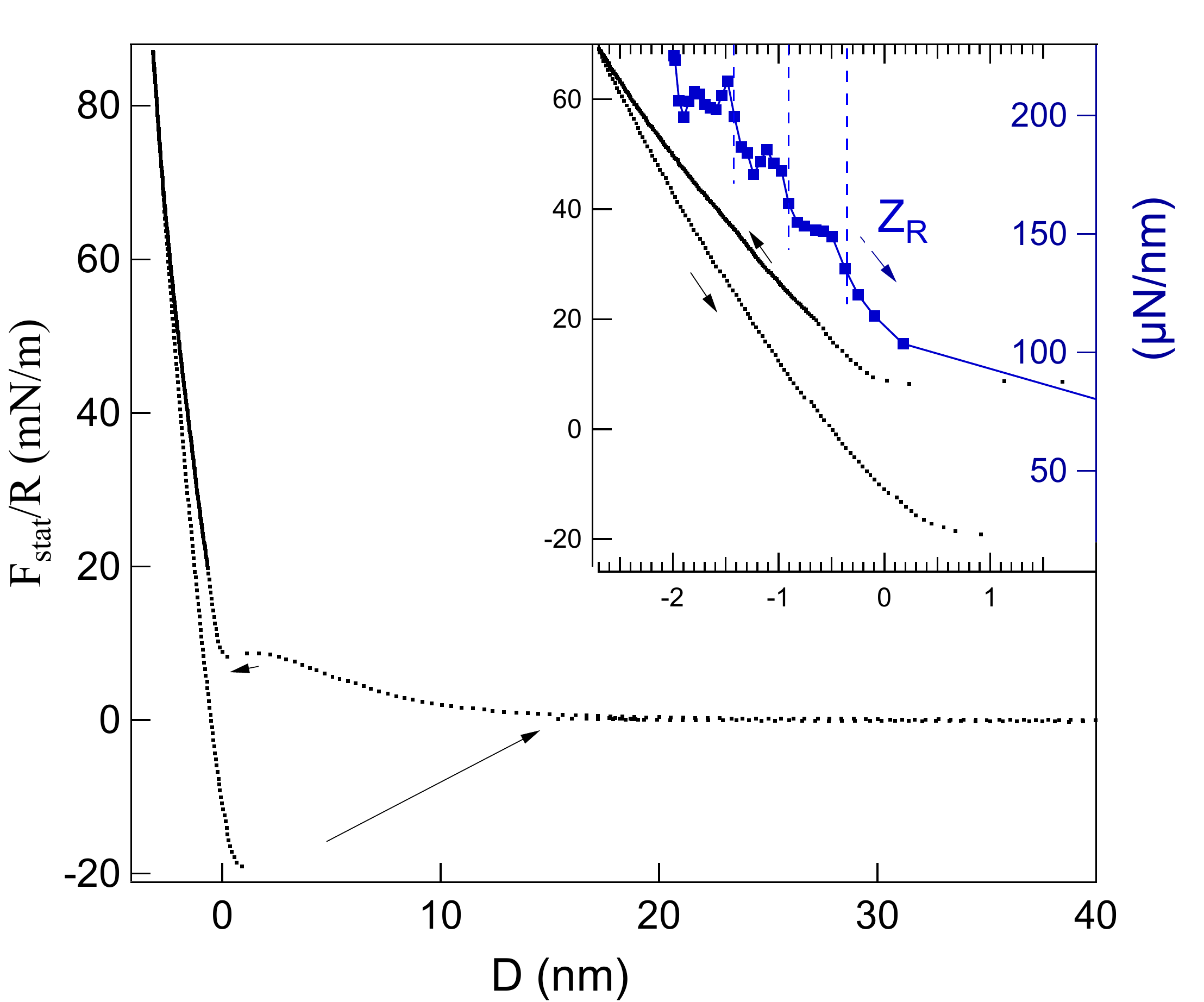}}
\caption{Reduced equilibrium force $F_{\rm stat}/R$ measured in a HPAM solution at 0.8 g/L concentration, as a fonction of the distance between the surfaces. The distance origin is set to the reception of the jump-to-contact of the advancing force (small arrow). Inset: enlargement close to contact. The right axis of the inset plots the dynamic stiffness $Z_R$ in $\mu$N/nm, measured at 220 Hz during the receding motion of the surfaces.   }
\label{fig:III1}       
\end{figure}


After a first cycle of approach, contact and withdrawal of the surfaces, which is singular, the subsequents cycles are highly reproducibles. This first cycle irreversibility has already been observed in SFA studies of various polyelectrolytes confined between oppositely charged or neutral surfaces \cite{Claesson2005}. In the following we present and discuss only the reproducible cycles after the first one. 

Figure (\ref{fig:III1}) plots the typical interaction force normalized by the sphere radius $R$ $F_{\rm stat}/R$ measured in HPAM solutions. Upon approaching the surfaces a repulsion interaction starts at a distance of around 20 nm from contact, followed by a jump-to-contact. We define the origin of the sphere-plane distance at the reception of this jump. However this is not a platinum-platinum contact: the stiffness $Z_R$ inside the contact increases by discrete steps, showing the expulsion of  polyelectrolyte layers. The size of the steps can be estimated to 5.2 \r{A}, which is close but slightly larger than the monomer size $a \approx 4 $\r{A}. Therefore the adsorbed layer is a dense, compact layer, as expected and observed for the adsorption of polyelectrolytes from a low ionic strength solution \cite{Claesson2005,DobryninRubinstein2005}. On separating the surfaces a	significant adhesion is measured, corresponding to a $F/R$ ratio of 19 mN/m, and to an interfacial tension $\gamma_{SL}=F_{\rm stat}/4\pi R = 1.6$ mN/m. It is usually considered that the adhesion force between polyelectrolyte adsorbed layers increases with the charge density of the polyelectrolyte, and $F/R$ ratios as high as 100-200 mN/m were observed for a charge fraction $f=1$ on mica \cite{Claesson2005}. The magnitude found here for HPAM of charge fraction f=0.25 is in good coherence with this tendancy. 

\begin{figure}[!ht]
\center\resizebox{0.45\textwidth}{!}{%
  \includegraphics{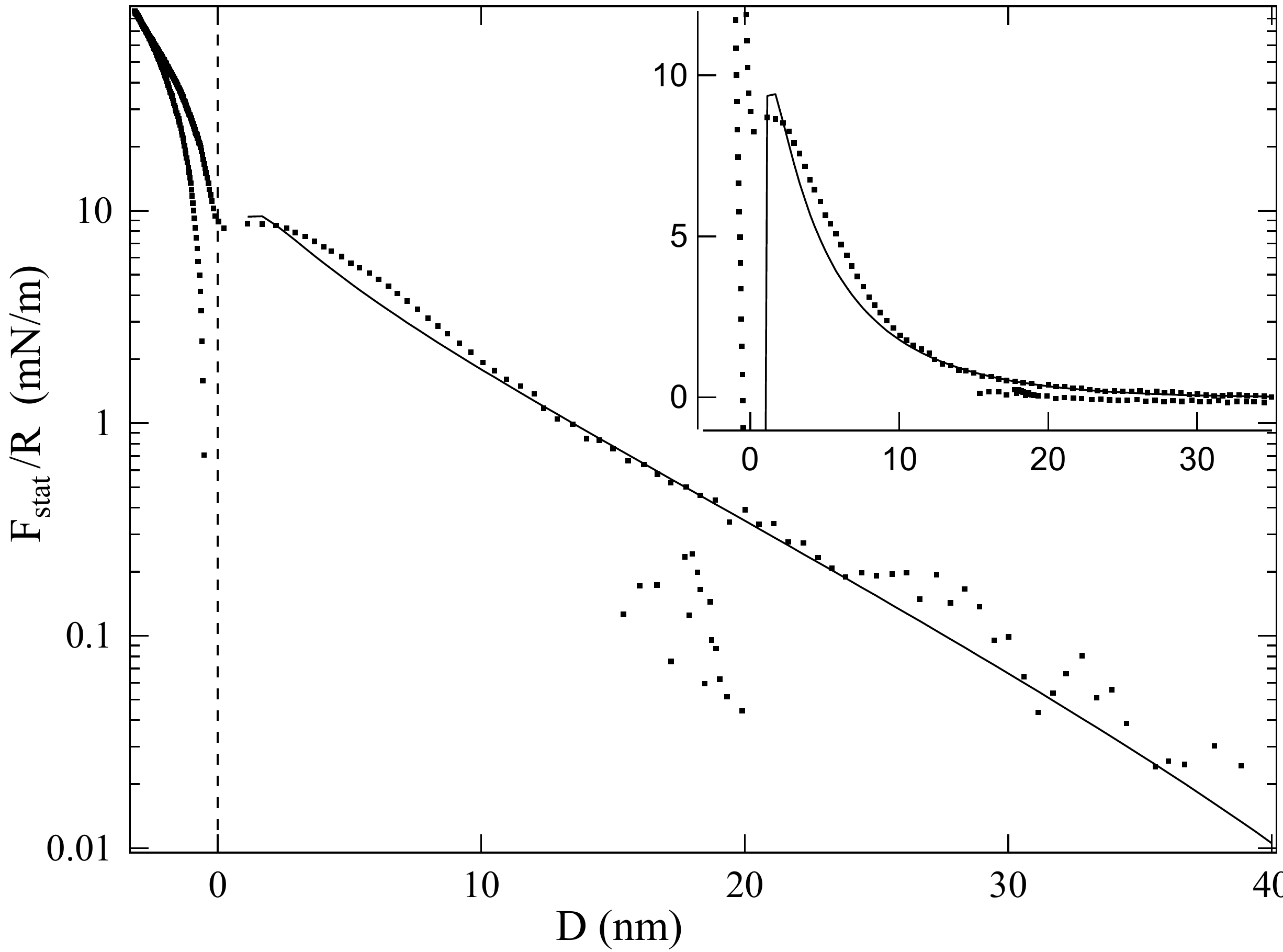}}
\caption{Dots: reduced equilibrium force $F_{\rm stat}/R$ plotted in log-linear scale. Continuous line: DLVO force at constant surface charge of 31 mC/m$^2$ with a Debye's length $l_D=5.8$ nm and a Hamaker constant 3.10$^{-20}$J.  The inset is an enlargement in linear scale. }
\label{fig:III2}       
\end{figure}

As shown on figure (\ref{fig:III2}) the repulsion interaction on approaching the surfaces is comparable in amplitude  to a DLVO interaction. In fig. \ref{fig:III2} ($C=0.8$ g/l)  the Debye's length is 5.8 nm and the surface  charge is 31 mC/m$^2$. Double Layer force were previously measured in polyelectrolyte solutions with the SFA \cite{Pincus2002}. We have used here a DLVO calculation for a monovalent electrolyte, although there is no added salt in the solution, and the polyelectrolyte itself is not exactly a monovalent electrolyte. We justify this choice by the fact that the electrical double layer neighboring the charged surfaces is essentially made of monovalent counterions extracted from the bulk electrolyte reservoir. At the experimental temperature of 28 $^o$ C with the relative permittivity of water of 77.22, the Debye's length of a monovalent electrolyte  solution is $l_D=0.304/\sqrt{c}$ nm with $c$ in mol/l, giving an equivalent bulk concentration of  monovalent electrolyte of 2.75 mM/l for $l_D$=5.8 nm. With the monomer mass
$M_A=71.25$ g and the ionization ratio $f\simeq 0.25$, the solution of 0.8 g/l corresponds to a monovalent charge concentration of 2.77 mM/l, which is extremely close to the value deduced from the Debye's length. This supports the interpretation in terms of DLVO interaction and the electrostatic nature of the repulsive interaction force. 

Thus, together with the stepped profile of the stiffness inside the contact, the picture emerging from the equilibrium interaction force is a solid/solution interface made of chains adsorbed in a compact layer and  creating a negative surface charge. The negatively charged surface is neighbored by an Electrical Double Layer (EDL) made of positive monovalent ions,  from which the positively charged free chains are essentially excluded.  
Polyelectrolyte adsorption on surface has been the focus of extensive studies, due in part to the technological importance of polyelectrolyte multilayers formed by the successive deposition of
positively and negatively charged polyelectrolytes from aqueous solutions \cite{DobryninRubinstein2005}.
In the case of a low salt concentration, which is the case here, the overcharge of the surface due to the chains adsorption is expected to be given by \cite{DobryninRubinstein2005}:
\begin{equation}
\delta \sigma =  e\frac{f^{1/3}}{u^{1/3}al_D} \qquad u=\frac{l_B}{a}
\end{equation}
With the experimental value of the Debye's length $l_D=5.8$ nm the above expression gives a surface charge of 35 mC/m$^2$. This theoretical value is very close to the experimental value issued from the DLVO fit of the repulsive force. A schematic picture of the interface picture is drawn in inset of figure (\ref{fig:V1}).


\section{Dynamic properties of confined polyelectrolyte solutions}
\label{sec:dynamic}
\subsection{Thick films}
\label{sec:thickfilms}

\begin{figure}[!t]
\resizebox{0.5\textwidth}{!}{%
  \includegraphics{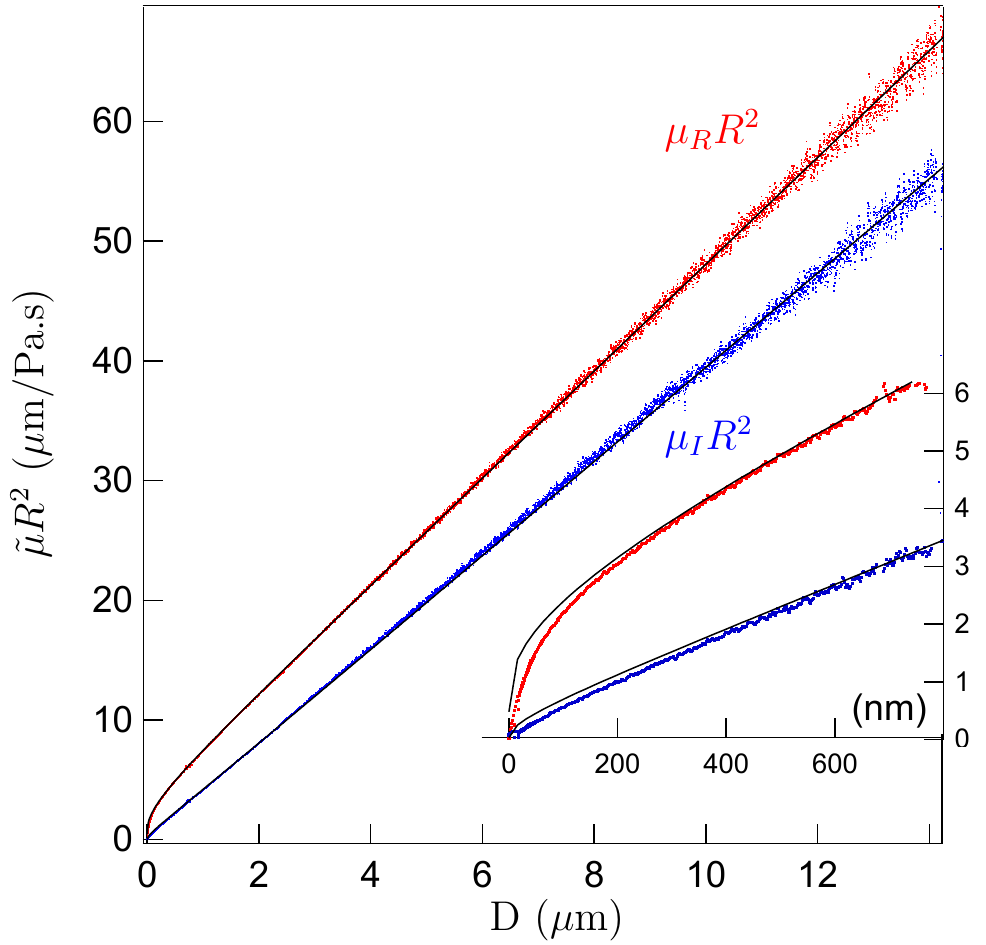}}
\caption{Components of the reduced mobility $\tilde \mu R^2 $  measured in an HPAM solution at a frequency of 220 Hz. Red dots: dissipative part $\mu_R R^2$, blue dots: elastic part $\mu_I R^2$.  The black lines are the components of the extended Hocking expression (\ref{Hocking}) fitted with a real-valued boundary friction coefficient $\tilde \lambda=\lambda_R$. Inset: enlargement below the micrometric scale. 
}
\label{fig:IV1}       
\end{figure}

Figure (\ref{fig:IV1}) plots the typical variation of the components of the reduced mobility measured in a HPAM solution as a fonction of the sphere-plane distance. At distances larger than the micrometer,  the variation of both components is essentially a straight line, as expected for a visco-elastic fluid.  From eq. (\ref{far_field}) the slope of these straight lines reflects the components of the complex visco-elastic modulus of the solutions. Both components have the same order of magnitude, which demonstrates the strong visco-elastic character of the solutions. The plot also shows that the flow does not  obey a no-slip boundary condition on the solid surfaces: the large distance variation of the dissipative part $\mu_R R^2 $ is not linear but affine with the distance. It extrapolates to a finite ordinate at origin, pointing toward values of the solid-liquid coefficient in the range of 30 kPa.s/m. To the contrary the conservative part $\mu_I R^2 $  extrapolates toward origin within the experimental uncertainty, which suggests a purely dissipative friction at the solid-liquid interface.

In order to determine precise values of the bulk modulii and  partial slip boundary condition,
 we compare the data to the exact theory given by eq. \ref{Hocking} for the oscillating drainage flow  of a visco-elastic liquid undergoing a Navier boundary condition. According to the experimental observation, we assume a purely dissipative friction coefficient $\tilde \lambda = \lambda_R$. The  quantitative agreement with the data at distances larger than 
200 nm is excellent (see inset of fig. \ref{fig:IV1}). It allows one to conclude that at large distance, the flow is indeed fully described by an apparent slip condition.  

\begin{figure}[!ht]
\resizebox{0.45\textwidth}{!}{%
  \includegraphics{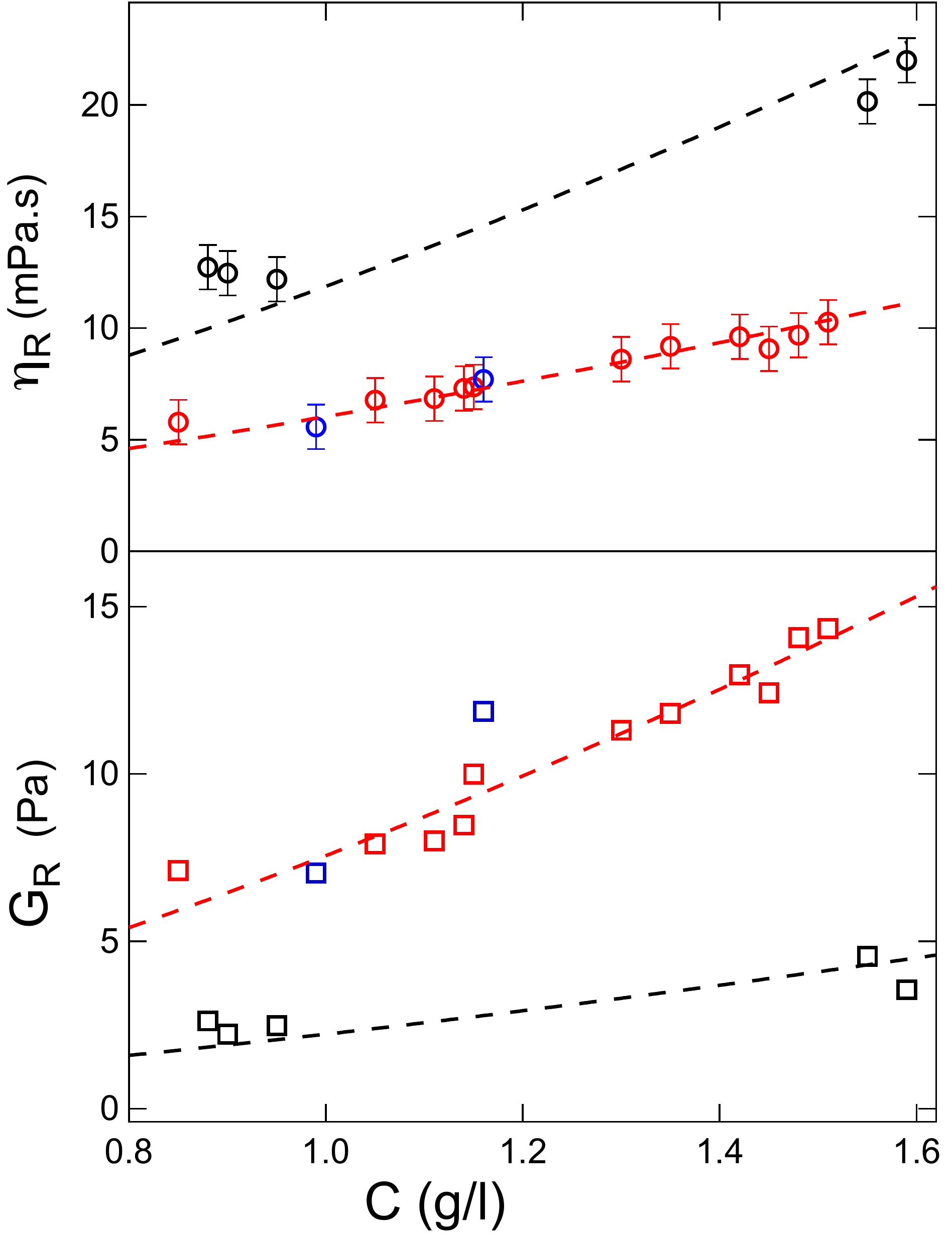}}
\caption{Top graph: real value of the viscosity $\eta_R$ as a function of the HPAM concentration 
($\circ$) 30 Hz, (\textcolor {red}{$\circ$}) 220 Hz, (\textcolor {blue}{$\circ$}) 248 Hz.
The dashed-lines are best power-law fits in $c^{3/2}$, $\eta_R=\eta_s+Ac^{3/2}$.  The prefactor $A$ depends on frequency. Bottom graph: real value of the shear modulus $G_R$ at different frequencies  ($\square$) 30 Hz, (\textcolor {red}{$\square$}) 220 Hz, (\textcolor {blue}{$\square$}) 248 Hz. The dashed-lines are best power-law fits in $c^{3/2}$, $G_R=Ac^{3/2}$.}
\label{fig:IV2}       
\end{figure}

The values of the visco-elastic modulii of the HPAM solutions measured for concentrations between 0.8 and 1.6 g/l  are plotted in fig. \ref{fig:IV2}. Both the viscosity $\eta_R$ and the shear modulus $G_R$ depend on the concentration and on the experimental frequency. For the variation with the concentration, we find a good agreement with a power law in exponent 3/2 (dashed lines in fig. \ref{fig:IV2}). This is in good agreement with the theoretical expectations for polymer solutions in the semi-dilute entangled regime \cite{DobryninRubinstein2005}, which is the case of our solutions. More precisely, the steady-stage shear viscosity and the plateau shear modulus of semi-dilute entangled polyelectrolyte solutions are expected to scale as $c^{3/2}$. We observe that these scaling laws hold for the  linear visco-elastic modulus at finite frequency measured here. For the frequency variation,  the ratio of the prefactors $A$ of the $c^{3/2}$ power law followed by the viscosity $\eta_R$,  $A(220)/A(30)=2.15=(220/30)^{-0.38}$,  is compatible with a variation  $\eta_R \propto \omega ^{-n}$ with $1/3 \leq n \leq 1/2$. The similar ratio for the shear modulus $G_R$,  $A(220)/A(30)=3.4=(220/30)^{0.61}$, is compatible with a variation $G_R \propto \omega ^{n}$ with $1/2 \leq n \leq 2/3$.



\begin{figure}[!t]
\resizebox{0.45\textwidth}{!}{%
  \includegraphics{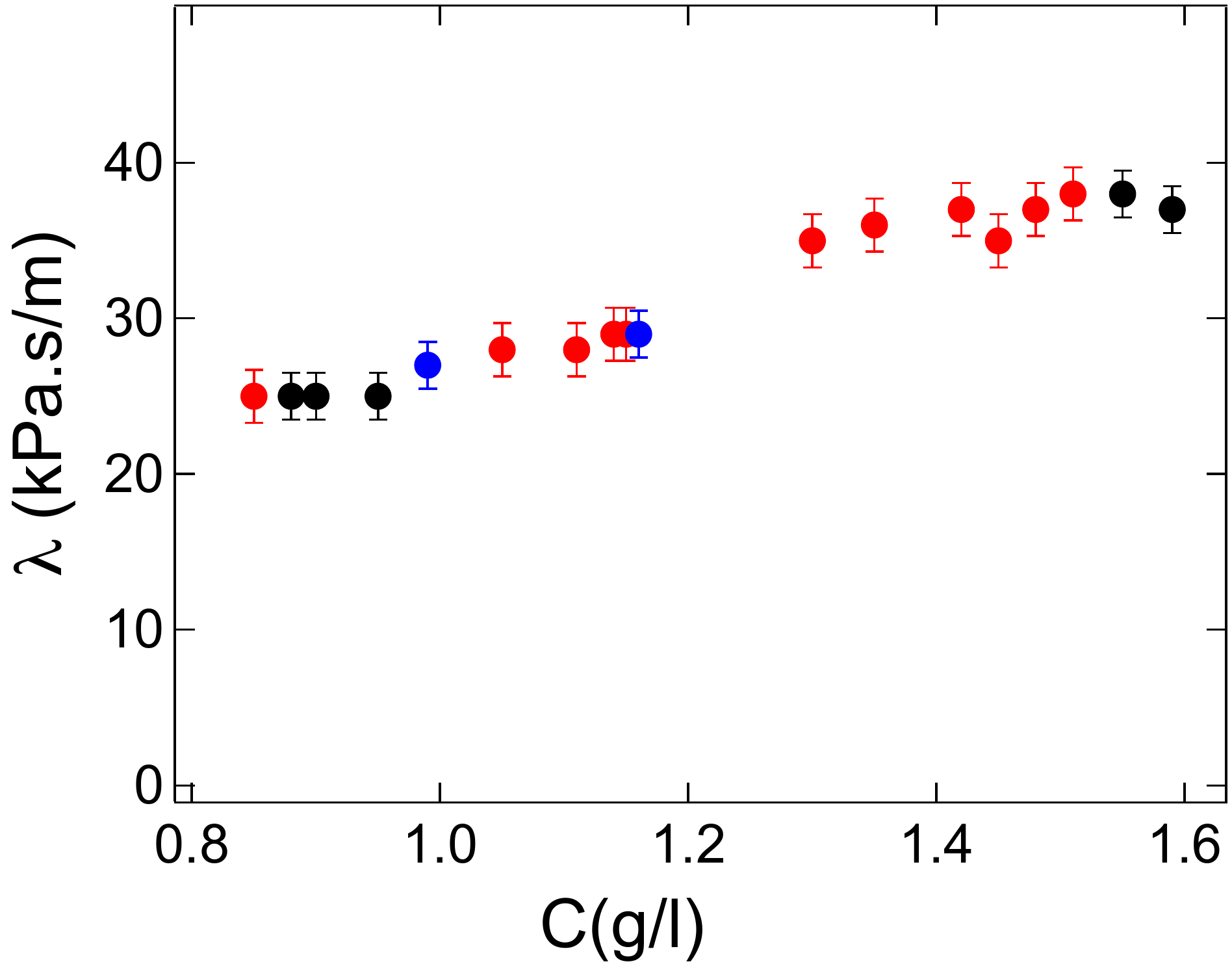}}
\caption{Friction coefficient at the HPAM solutions/solid interface as a function of the concentration at different frequencies:
30Hz ({\Large${\bullet}$}), 220 Hz, (\textcolor {red}{\Large${\bullet}$}) 248 Hz, (\textcolor {blue}{\Large${\bullet}$}) 248 Hz.
}
\label{fig:IV3}       
\end{figure}

The values of the boundary friction coefficient $\tilde \lambda = \lambda_R$ of the solutions are plotted in fig. \ref{fig:IV3}. The  boundary friction is not only purely dissipative, but also independent of frequency in the range studied. This should be emphasized, considering the significant variation with frequency  of the bulk modulus of the solution. In contrast to the bulk visco-elastic character of the polyelectrolyte solutions, the friction coefficient on the wall appears to be essentially Newtonian,  real-valued and frequency independent. 

The absolute value of the slip lengths $\vert \tilde \eta \vert/\lambda $ associated to the interfacial  friction coefficients found here lie in the range of 250 nm (248 Hz) to 850 nm (30 Hz). This is significantly smaller than the slip lengths of 5 $\mu$m to 15 $\mu$m obtained by Cuenca et al \cite{cuenca_submicron_2013} in submicrometric and micrometric nanochannels. However, this apparent discrepancy might be a spurious effect of using the slip length to characterize the boundary condition of a non-Newtonian liquid.  As shown in \cite{cross2018} the slip length does not truly  reflect the interfacial boundary rheology of a complex fluids, as it incorporates the properties and variations of its bulk  rheology. The bulk viscosity of HPAM solutions in Cuenca et al is a steady-state viscosity, of values in the range 0.01 - 0.1 mPa.s,
 somewhat larger than our finite-frequency viscosities. The ratio $\eta/b$ in their experiments lies in the tens of kPa.s/m, slightly lower but of similar magnitude than the friction coefficients measured in our experiment.  
The larger slip length reported by Cuenca et al could just be an effect of the bulk viscosity decay with frequency in the non-Newtonian HPAM solutions,   not related to the boundary hydrodynamics. 

In view of the picture emerging from the equilibrium force of a liquid/solid interface  containing adsorbed chains and an EDL made of counterions, it is tempting to attribute the  fully Newtonian friction of the solutions onto the solid wall to the lubrication effect of  a thin layer of low viscosity liquid.  Large wall slippage of polyelectrolyte solutions have indeed been
reported previously in porous media, membranes \cite{chauveteau_concentration_1984,Chauveteau1,chauveteau2},
Surface Force measurements \cite{cayer-barrioz_drainage_2008}, as well as in solid-state
nano-fluidic channels \cite{cuenca_submicron_2013}, and attributed to a depletion layer at the solid/liquid interface. Assuming a depletion layer of pure  water of viscosity  0.85 mPa.s at the experiment temperature, the friction coefficients of fig. \ref{fig:IV3} would correspond to values of the depletion layer thickness $e_d\approx \eta_{water}/\lambda_R$  between 40 nm at 0.8 g/l to 26 nm at 1.6 g/l. 
The next paragraph analyzes in more detail the mobility of polyelectrolyte solutions confined at this scale.

\subsection{Thin films}

\begin{figure}[!t]
\resizebox{0.49\textwidth}{!}{%
  \includegraphics{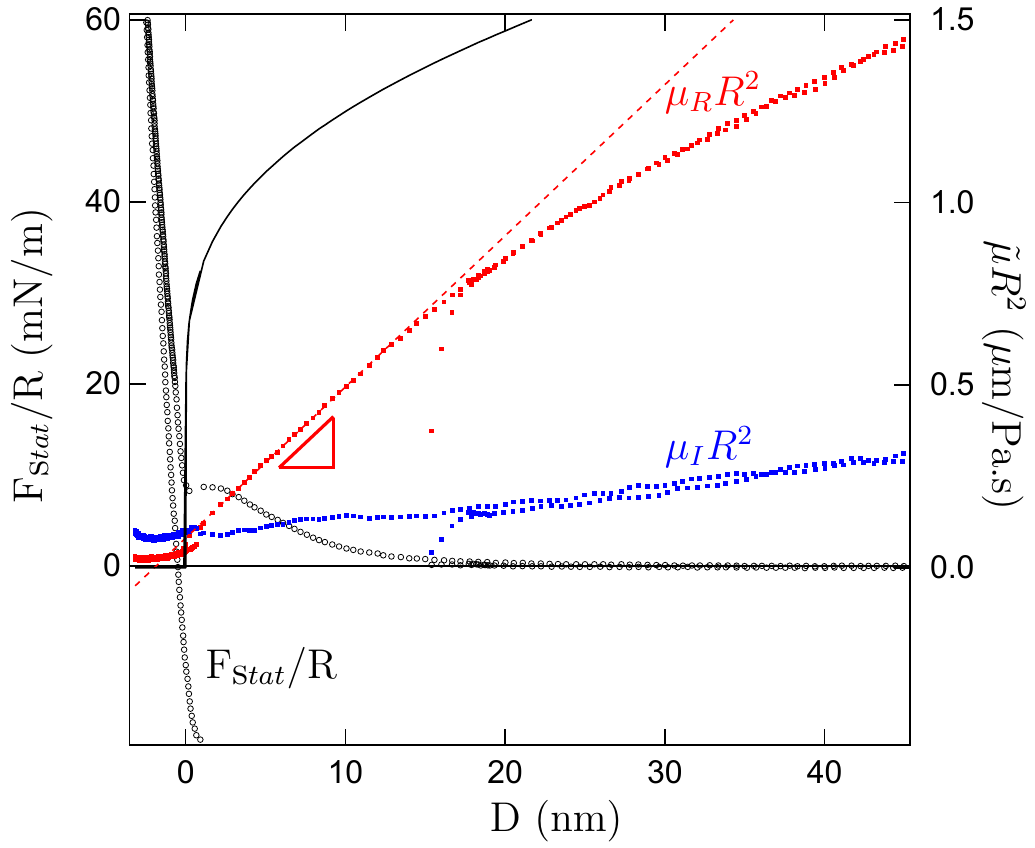}}
\caption{Left axis ($\circ$): reduced equilibrium force $F_{stat}/R$ measured in HPAM solutions at g/l. Right axis: components of the reduced mobility $\mu_R R^2$ (red dots) and $\mu_I R^2$ (blue dots) measured simultaneously at 220 Hz. The black continuous line is the best fit with the model of partial slip at wall (eq. \ref{Hocking}). The red dashed line is the reduced mobility of a sphere-plane gap filled with a liquid of viscosity $\eta_1=0.9 mPa.s$. 
}
\label{fig:IV4}       
\end{figure}

The dynamic components of the reduced mobility in a typical experiment, at  sphere-plane distances below 40nm,  are plotted on fig. (\ref{fig:IV4}) superimposed with the static force $F_{\rm stat}/R$. 
It appears clearly  at this scale, that the prediction of the flow model with a partial slip boundary condition located at the distance origin, overestimates badly the dissipative component of the mobility, by a factor as large as 4. More specifically, in the range of distance where a significant repulsive interaction force is measured, the dissipative component of the reduced mobility increases linearly with the distance, in a manner equivalent to that of a liquid of viscosity $\eta_1$ flowing with a no-slip or very small slip b.c. (less than 2 nm) with respect to the chosen distance origin. The viscosity $\eta_1$ as extracted from the slope is of $0.9 $ mPa.s in this specific experiment, very close to the viscosity of water in these conditions. 

This behavior supports clearly the existence of a layer of reduced viscosity at the adsorbed polyelectrolyte/solution interface. A decrease of the average viscosity of thin polyelectrolyte films has already been observed in pionnering dynamic SFA experiments \cite{Tadmor-Israelachvili2002,Kuhl-Israelachvili1998}. In hyaluronic acid confined between non-adsorbing mica surfaces, Tadmor et al \cite{Tadmor-Israelachvili2002} found a linear decrease of the average viscosity below a gap of 400 nm, reaching the pure water viscosity at zero film thickness. Kuhl et al \cite{Kuhl-Israelachvili1998} found a similar reduction of the average viscosity of aqueous polyethyelne glycol films confined between bilayers. In both cases, the hydrodynamic force was interpretated in terms of effective (average) liquid viscosity for a given value of the confinement, and the limit viscosity was found to be that of the solvant. 

\begin{figure}[htb]
\resizebox{0.45\textwidth}{!}{%
  \includegraphics{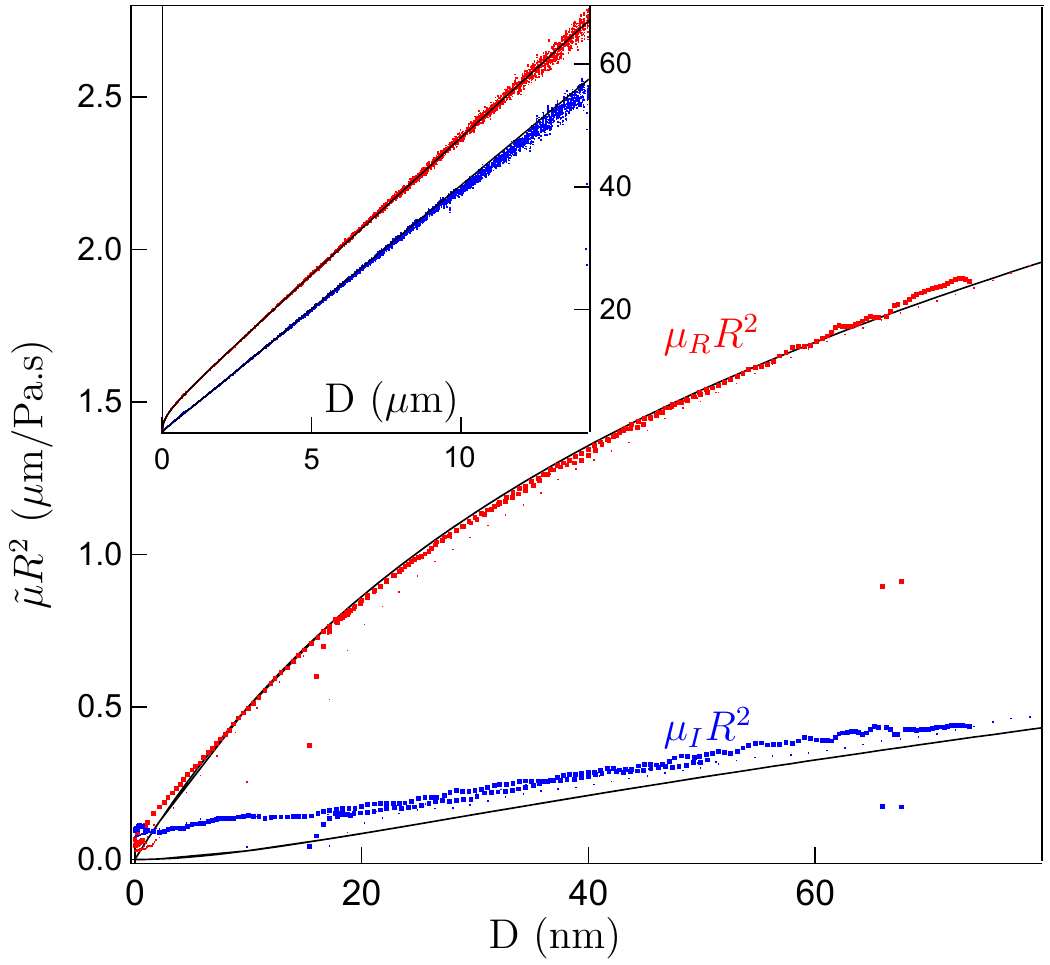}}
  \resizebox{0.45\textwidth}{!}{%
  \includegraphics{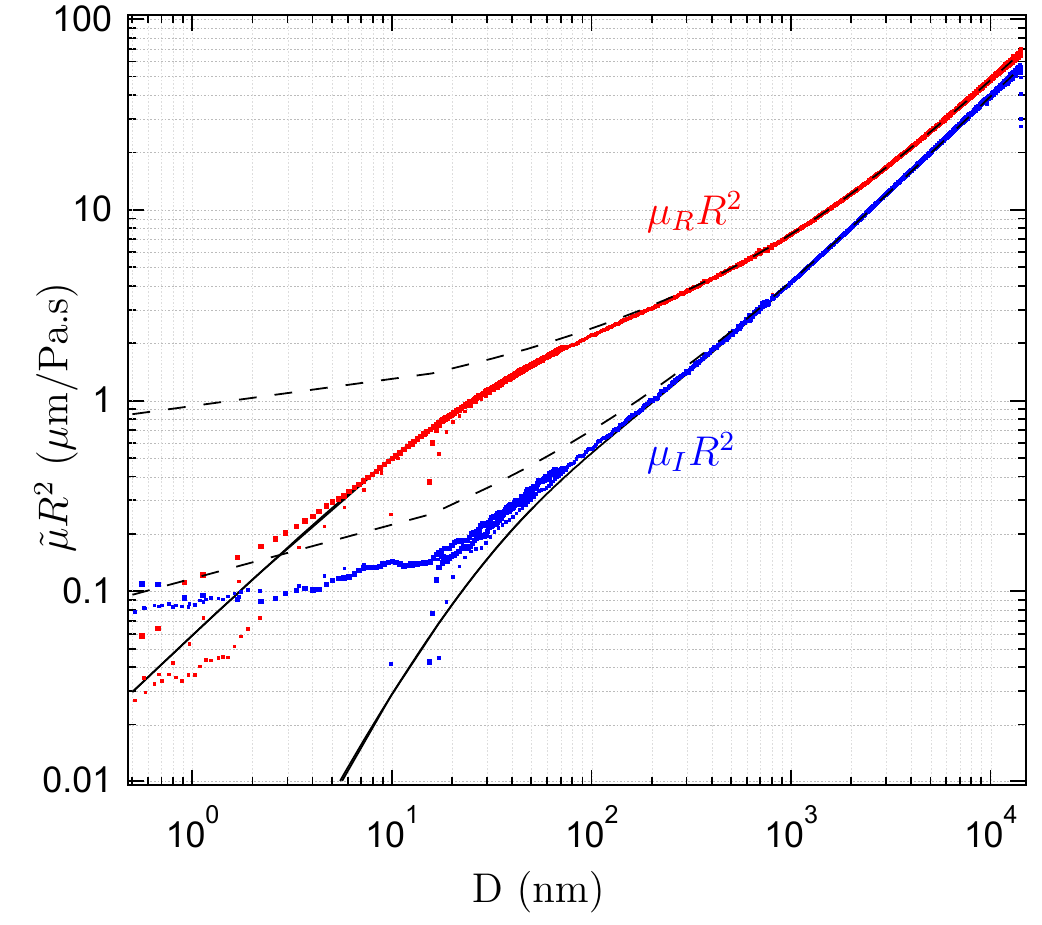}}
\caption{ Components of the reduced mobility in a HPAM solution of 1 g/l at 220 Hz (red: dissipative component, blue: elastic component) as a function of the sphere-plane distance. (a) Linear scale. The black continuous lines are the components of the two-fluid model (eqs \ref{2fluidmodel}) calculated with $e=33$ nm and $\eta_1=0.85$ mPa.s. The inset shows the plot at the micrometric scale. (b) log-log scale. The black dashed-line is the theory for a partial slip boundary condition enforced at D=0 (eq. \ref{Navier}).
}
\label{fig:IV5}       
\end{figure}

In order to determine more precisely the thickness of the reduced viscosity layer in our experiments,  we compare our data with a two-fluid flow model,  incorporating on each surface one layer of thickness $e$  and of viscosity $\tilde \eta_1$, and a sharp viscosity contrast with the bulk
(of visco-elasticity $\tilde \eta$). The reduced mobility writes (see Appendix): 
\begin{eqnarray}
\tilde \mu R^2= \frac{D}{6 \pi \tilde \eta}\left (  f^*\left (\frac{D}{e}\right ) \right )^{-1}\qquad \tilde \delta = \frac{\tilde \eta}{\tilde \eta_1}-1\cr
f^*(x\ge 2) = 2x \sum_{n=1}^3 \alpha_n(x-A_n)\ln(1-A_n/x)\cr
f^*(x\le 2) =  2x \sum_{n=1}^3 \alpha_n(x-A_n)\ln(1-\frac{A_n}{2})+\frac{(x-2)^2}{4(\delta+1)}\cr
  \alpha_n=\frac{1}{12\delta(\delta+1)\gamma_n} \qquad A_n=-2\tilde \delta j^n\beta\gamma_n \cr
 \gamma_n=\frac{1}{j^n\beta}+1+j^n\beta \qquad (j^n\beta)^3=\frac{1+\tilde \delta}{\tilde \delta}
 \quad.
 \label{2fluidmodel}
\end{eqnarray}
At small values of the distance $D\ll e$, the mobility reduces to $\tilde \mu R^2 \simeq D/6 \pi \eta_1$ (see Appendix). Therefore the actual viscosity of the interfacial layers can  be determined from the slope of the components of $\tilde \mu R^2$ at small distance.

We compare the two-fluid model to the experimental data by keeping the value of the bulk visco-elastic modulus $\tilde \eta$ equal to that found in the thick film analysis. We also enforce a viscosity $\eta_1$ of the fluid layer coating the surfaces, equal to that of water, in good coherence with the slope of the data at small distances. Thus only the thickness $e$ of the 
coating layers is adjusted. Figure \ref{fig:IV5} shows that the two-fluid model provides a much better description of the mobility at the microscopic scale, than the partial slip theory enforced at $D=0$. In the exemple of the figure (concentration 1 g/l, frequency 220 Hz) the layers thickness found  is $e=33$ nm. The two-fluid model is essentially identical to the partial slip boundary condition model at distances $D \ge 10 e$. At smaller scale (larger confinement) the finite thickness of the lubricating layer at the wall interface cannot be ignored anymore, and the partial slip b.c. model becomes inaccurate. The two-fluid model is in excellent agreement with the dissipative part of the mobility from the macroscopic scale down to 4 nm, for a unique value of the boundary layer thickness $e$. The deviation with the data below 4nm, is compatible with expected elasto-hydrodynamic effects due to the elastic deformation of the confining surfaces, as described in \cite{LeroyJFM2011,VilleyPRL2013}. The deviation of the two-fluid model from the elastic component of the data at small scale extends to a somewhat larger distance. The slightly larger value of the data, could be due to the effect of a fluid surface tension between the two layers, not taken into account in the model.

\begin{figure}[!h]
\resizebox{0.42\textwidth}{!}{%
\includegraphics{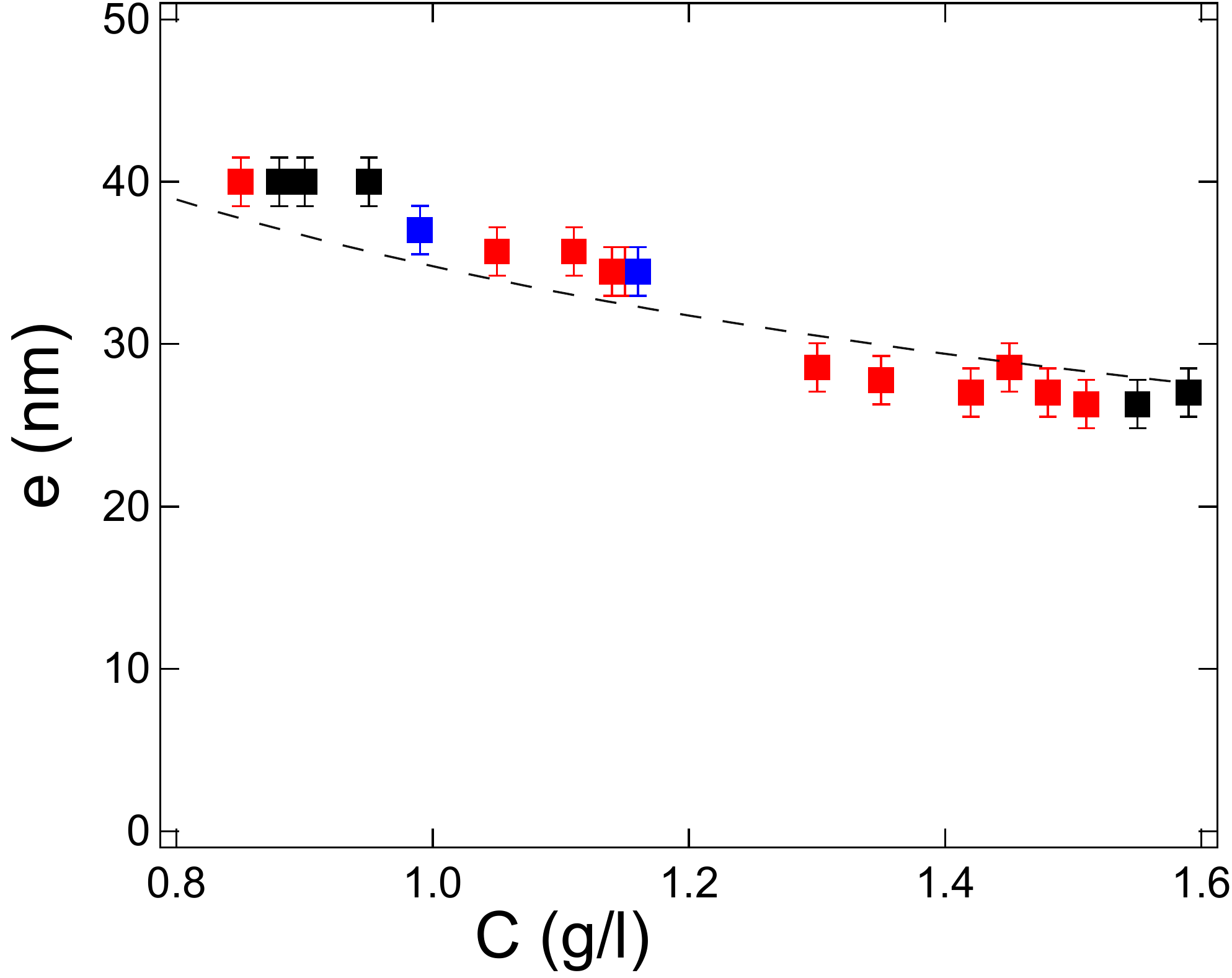}
}
\caption{Thickness of the interfacial layer of water $e$ obtained from the two-fluid model, as a function of the concentration of the solution C in g/l: ($\blacksquare$) 30 Hz, (\textcolor {red}{$\blacksquare$}) 220 Hz, (\textcolor {blue}{$\blacksquare$}) 248 Hz. The dashed black lines  plots the power-laws  $y=(\xi+2l_D) C^{-1/2}$  with $\xi=23 $ nm and $l_D = 5.8$ nm. }  
\label{fig:IV6}    
\end{figure}

Figure (\ref{fig:IV6}) shows the thickness values $e$ of the low viscosity interfacial layers,  obtained from the fit of the data with the two-fluid model. For all the analyzed data, the value at small distance of the slope of the dissipative component $\tilde \mu_R R^2$ is well described by the viscosity of pure water (as in fig. \ref{fig:IV4}) therefore the interfacial viscosity $\eta_1$ is kept equal to $\eta_{water}$. The values of the thickness  for concentration ranging from 0.8 to 1.6 g/l, vary between   26 nm to 40 nm, in good quantitative agreement with what can be  expected from the  friction coefficient  of the thick film model $e \simeq \lambda \eta_{\rm water}$. 
One also observes that as $\lambda$,  the thickness $e$ does not depend on the forcing frequency. 


The thickness of the interfacial layer depends significantly on the concentration of the semi-dilute solution, decreasing by a ratio of 1.5 when the concentration increases by a factor of 2. In the limited range of concentration studied, the variation of the thickness $e$ is compatible with a power law $c^{-n}$ with $1/2 \le n \le 3/4$. The figure \ref{fig:IV6} shows a comparison with a power law in $c^{-1/2}$.

\section{Discussion}
\label{sec:discussion}
Our nanorheology experiments performed on the dynamic Surface Force Apparatus  confirm the 
low friction  of poly-electrolyte solutions on solid surfaces, as already observed in flow experiments involving other type of confined geometries (porous media, membranes \cite{chauveteau_concentration_1984,Chauveteau1,chauveteau2}, solid-state
nano-channels \cite{cuenca_submicron_2013} and SFA 
\cite{cayer-barrioz_drainage_2008}).

In these previous experiments, the low friction at the solution/solid interface was interpreted in terms of a large slip length. Our results  strongly  suggest that the value of the slip length is not appropriate to compare the wall slip of non-Newtonian liquids, as the actual relation between the shear stress and the shear rate at the wall may depend on the experimental conditions. The values of the slip lengths found at the micrometric scale in the present oscillatory experiments are more than one order of magnitude smaller than that found by Cuenca et al in steady-state conditions for similar confinements. However the friction coefficients in the two sets of experiments are of same magnitude, in the tens of kPa.s/m.
Investigating the boundary conditions of the solutions over one decade of frequency, we find that the solid/solution friction coefficient is fully Newtonian: the slip velocity follows  the shear stress at the wall instantaneously and independently on the frequency.

The extended spatial range of our dynamic Surface Force Apparatus allows one 
to clearly demonstrate that the low friction of the solution at the wall   is associated to the presence of a low viscosity  fluid layer  coating the solid surfaces, whose thickness and viscosity are directly resolved. Incorporating this low viscosity layer in a classical two-fluids hydrodynamic model, we find that the hydrodynamic force is described accurately from the macroscopic scale down to 4 nanometers. This result does not support the finding of Cuenca et al of a wall slip reduction with increasing confinement \cite{cuenca_submicron_2013}. However, the two-fluid model shows that the partial slip boundary condition is a macroscopic approximation which holds only at confining distances ten times larger than the thickness of the lubricating layers. At smaller gaps, the partial b.c. model 
overestimates the liquid mobility. 
Therefore, as Cuenca et al estimate the slip b.c. from flow rate measurements in nanochannels of height less than ten times the thickness of their depletion layer, we think that the reduction of slip that they report with increased confinement, may be partially explained by the too severe approximation of neglecting the finite thickness of the depletion layer.

We discuss now the physical origin of the boundary low viscosity layer. A strong characteristic revealed by the present experiments, is   that  the layer thickness does neither depend on the frequency nor on the sphere-plane distance, and is thus independant on the flow shear rate. More specifically, by combining the different frequencies investigated  with the variation of the sphere-plane distance, we estimate that the domain of shear rated covered in these dynamic force measurements lies between $10^{-5}$ to $10^{-2}$ s$^{-1}$. 
 The absence of shear rate dependency over this range does not support a dynamic mechanism of formation of the boundary layer, such as shear-induced migration of polymer chains, segregation under flow, or similar flow-induced structural changes at the interface,  which have been reported at higher shear rates \cite{graham2011}. Rather, it suggests that in our experimental conditions, the presence of the low viscosity  layer coating the solid surfaces is an equilibrium property of the solid/solution interface.

 \begin{figure}[!t]
\resizebox{0.5\textwidth}{!}{%
  \includegraphics{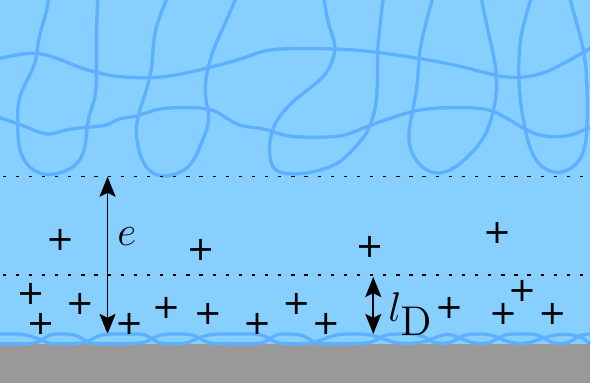} \includegraphics{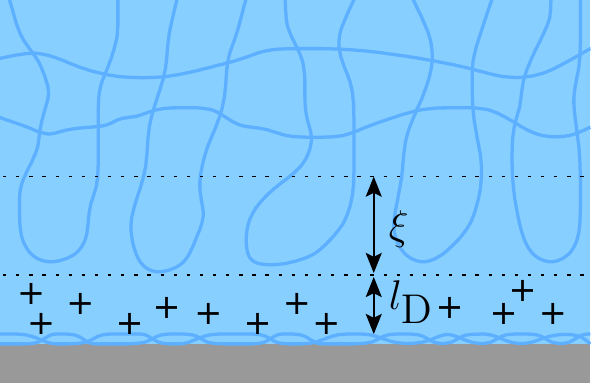}
  }
\caption{Schematic drawing of possible scenarii for the low viscosity layer at the solution/solid interface. }
\label{fig:V1}    
\end{figure}

The values of the thickness $e$ of the boundary layer are significantly larger than  the Debye's length $l_D$ screening the electrostatic interactions with the solid wall and the chains adsorbed on it. It is also larger than the estimated correlation length $\xi$  of the semi-dilute solutions (23 nm at 1 g/l). But the variation of $e$ with the solution concentration  is compatible with that of these two quantities, which both vary as $C^{-1/2}$. Based on this 
considerations one can think of two scenarii for the microscopic origin of the boundary low viscosity layer.

In the first scenario the low viscosity layer  is an actual depletion layer, containing essentially no or very few polymer chains. The latter  are expelled out of the boundary layer by equilibrium interactions. However the equilibrium interactions between the surfaces are directly measured by the SFA, and they appear to be negligible at a distance $D=2e$ where the depletion layers start to overlap. The reason why the polymer chains  would be repelled away from the surface on a scale significantly larger than the scale of the interaction, here the Debye's length $l_D$, is unclear.

In the second scenario   the low viscosity layer is not a depletion layer. The concentration of the polymer in this layer is essentially  the same as in the bulk, except at a distance from the wall governed by the equilibrium interaction.   However, below the bulk correlation length of the semi-dilute solution, the bulk viscosity  is not achieved, as there are no significant chain interactions or entanglements.  The low viscosity layer  thus contains loose chain loops not interacting between themselves, but also not adhering to the solid surface from which they are repelled by the negative electrostatic potential. Therefore the effective viscosity in this layer could be close to that of the solvant. In this scenario, sketched in fig. (\ref{fig:V1}), the thickness of the low viscosity layer is $e\simeq \xi +\alpha l_D$ with $\alpha $ of order of unity.  A value $\alpha = 2$ could account for the measured value of the layer thickness, as plotted in fig. (\ref{fig:IV6}).

\section{Conclusion}
\label{sec:conclusion}

Thanks to a Dynamic Surface Force Apparatus we have measured in the same experiment
the equilibrium interactions and the hydrodynamic properties in a confined  polyelectrolyte solution. 
The polyelectrolyte chains adsorb on the surface in a dense compact layer, on which
the viscoelastic solution flows with an apparent slip boundary condition. The slip is described at large scale by a Newtonian interfacial friction coefficient, whose origin lies in a  layer of same viscosity as water,  separating the compact adsorbed layer and the bulk visco-elastic solution. 
The thickness of this low viscosity boundary  layer is directly resolved for the first time
to our knowledge. It is found to be an equilibrium property of the adsorbed polyelectrolyte/bulk solution interface, independant on shear rate, flow frequency, and confinement. 
It is slightly larger than the estimated correlation length of the semi-dilute solution. 

This work calls for additional theoretical work in order to
understand the relation between the local viscosity and the concentration profile
at a polyelectrolyte/wall interface. 
From an experimental standpoint, it would
be interesting to conduct structural experiments to determine the concentration profile
at the interface, and to extend both the concentration range,
the polymer properties, and the salinity of the solutions. 

%
%

%

\bigskip 
This research was supported by the ANR program
ANR-15-CE06-0005-02.

\bigskip

{\bf Appendix: two-fluid model}

\setlength{\unitlength}{0.9cm}
\begin{picture}(9,3.8)(0,-0.4)
\put(0,0){\vector(1,0){5}}
\put(4.8,-0.3){$r$}
\put(2.5,0){\vector(0,1){3}}
\put(2.2,3.0){\makebox(0,0){$y$}}
\put(4,0){\vector(0,1){1.7}}
\put(4,1.8){\vector(0,-1){1.7}}
\put(3.9,1.9){$z$}
\put(2.5,0){\vector(0,1){1.5}}
\put(2.5,1.5){\vector(0,-1){1.5}}
\put(2.7,1.7){$D$}
\put(0.0,0.4){$\eta_1 \ \ \ e$}
\put(0.5,0.6){\vector(0,-1){0.4}}
\put(0.5,-0.4){\vector(0,1){0.4}}
\put(0.1,1.5){\vector(1,2){0.2}}
\put(0.6,2.5){\vector(-1,-2){0.2}}
\put(0.7,2.1){$e$}
\put(4,3){\vector(1,-2){0.5}}
\put(4.1,3){$R$}
\put(1.5,0.7){$\eta_2$}
\thicklines
\qbezier(2.5,1.5)(4,1.5)(5,2.3)
\qbezier(2.5,1.5)(1,1.5)(0,2.3)
\put(0,0){\line(1,0){5}}
\thinlines
\qbezier(2.5,1.3)(4,1.3)(5,2.1)
\qbezier(2.5,1.3)(1,1.3)(0,2.1)
\put(0,0.2){\line(1,0){5}}
\linethickness{0.3mm}
\put(6,0){\line(1,0){2.6}}
\put(6,2){\line(1,0){2.6}}
\linethickness{0.075mm}
\put(6,0){\vector(0,1){3}}
\put(6.1,3){$y$}
\put(8.6,0){\vector(0,1){2}}
\put(8.6,2){\vector(0,-1){2}}
\put(8.4,2.1){$z$}
\put(6,0.3){\line(1,0){2.6}}
\put(6,1.7){\line(1,0){2.6}}
\thinlines
\put(6,0){\line(5,1){1.4}}
\put(6,2){\line(5,-1){1.4}}
\qbezier(7.8,1.0)(7.8,1.5)(7.4,1.7)
\qbezier(7.8,1.0)(7.8,0.5)(7.4,0.3)
\linethickness{0.15mm}
\put(6,0.3){\vector(1,0){1.4}}
\put(6,1.7){\vector(1,0){1.4}}
\put(6,0.8){\vector(1,0){1.7}}
\put(6,1.2){\vector(1,0){1.7}}
\put(7.8,0.5){$v_r(y)$}
\end{picture}
\bigskip

In this model the sphere and the plane are both covered with a liquid layer of thickness $e$ and of viscosity $\eta_1$. The viscosity of the surrounding liquid is $\eta_2$, and   $\delta = \eta_2/\eta_1-1$  is the relative excess viscosity.  At small gap $D\ll R$,  and if the sphere motion is slow  compared to the diffusion time across the fluid films ($\dot D \ll \eta_1/\rho D$, with $\rho$ the fluid density) the lubrication properties are met: the velocity profile $v_r(y)$ is parallel to the plane, the pressure $P(r)$ is uniform across the film thickness, and the average velocity is locally proportionnal to the pressure gradient:
\begin{equation}
u(r)=\frac{1}{z}\int_0^zv_r(y)dr = - \frac{K(z)}{\eta_2}\frac{dP}{dr} 
\label{35}
\end{equation}
The velocity profile obeys the Stokes equation in each phase, the no-slip b.c. on each solid surface, and the condition of continuity of velocity and tangential stress at the two liquid interfaces. One get:
\begin{eqnarray}
0\le y \le e \qquad  v_r(y)  = \frac{\nabla P}{2\eta_1}y(y-z)    \cr
 e \le y \le \frac{z}{2} \qquad v_r(y) = \frac{\nabla P}{2\eta_2}\left (y^2-z(y+\delta e)+\delta e^2 \right )\cr 
 \frac{z}{2} \le y \le z \qquad v_r(y) = v_r(z-y) \cr
z\ge 2e \qquad  K(z) =K_2(z) = \frac{z^2}{12}+\frac{e}{2}\delta \left ( z-2e+\frac{4e^2}{3z}\right ) \qquad   \cr
  z \le 2e \qquad K(z)=K_1(z)=\frac{z^2 \eta_2}{12\eta_1}\qquad .
  \label{permeability}
\end{eqnarray}
The volume conservation at velocity $\dot D$ writes: 
\begin{equation}
2\pi rzu(r)=-\pi r^2\dot D 
\end{equation}
Here we consider the total volume conservation only, and assume that the solvant exchange between the two phases ensures that the thickness $e$ remains uniform.This leads, using (\ref{35}) and the parabolic approximation $z=D+r^2/R$, to: 
\begin{equation}
P(z)=-\frac{R\eta_2 \dot D}{2}\int_z^\infty \frac{dz'}{z'K(z')}
\end{equation}
so that the hydrodynamic force writes:
\begin{equation}
F(D) = \int_0^\infty 2\pi rP(r) = - \pi R \eta_2 \dot D  \int_D^\infty dz \int_z^\infty \frac{dz'}{z'K(z')}
\label{force}
\end{equation}

\bigskip
In an oscillatory flow at frequency $\omega/2\pi$ the forcing velocity is $\dot D=$ Re[$ih_o \omega e^{i\omega t}$]. In the limit of linear response $h_o\ll D$, all the dynamic quantities are harmonic functions of time at the forcing frequency, and are characterized by their complex amplitude only. In the above equations, only the terms in first order of $h_o$ are retained. The set of equations (\ref{permeability}) remains valid with the viscosities $\eta_1$, $\eta_2$ replaced by their complex visco-elastic analogous $\tilde \eta_1$ and  $\tilde \eta_2$,  and eq. (\ref{force})  with $\eta_2$ replaced by the visco-elasticity $\tilde \eta_2$, gives the complex amplitude of the oscillating hydrodynamic force.

\bigskip

Defining the non-dimensional variable $\zeta = z/e$ and functions $\kappa_i(\zeta)=12zK_i(z)/e^3$, one get the following expressions:
\begin{eqnarray}
F(D) = -\frac{6\pi \eta_2 R^2 \dot D}{D} f^*(D/e) \cr
f^*(x) = 2x \int_x^\infty d\zeta \int_{\zeta}^\infty \frac{d\zeta '}{\kappa(\zeta ')} \cr
\zeta \le 2 \qquad \kappa (\zeta) = \kappa_1(\zeta) =(\delta+1) \zeta^3 \cr
.\cr
\zeta \ge 2 \qquad \kappa (\zeta) =\kappa_2(\zeta)=\zeta ^3 + 6 \delta \zeta^2 - 12 \delta \zeta + 8 \delta 
\end{eqnarray}

The 3 complex roots of $\kappa_2(\zeta)$ are:
\begin{equation}
A_n = -2 \delta (1+j^n \beta+j^{2n} \beta^2) \quad 
(j^n\beta)^3=\frac{1+\delta}{\delta} \quad j=e^{2i\pi/3}
\end{equation} 
(note that $\delta $ is a complex number) and its inverse  decomposes in algebraic fractions as:
\begin{eqnarray}
\frac{1}{\kappa_2(\zeta)} = \sum_{n=1}^3 \frac{\alpha_n}{\zeta-A_n} \qquad 
\alpha_n=\frac{1}{12\delta(\delta+1)\gamma_n}\cr
{\rm with} \quad \gamma_n=\frac{1}{j^n \beta}+1+j^n\beta  \cr
{\rm and} \quad \sum_{n=1}^3 \alpha_n=0 \quad \sum_{n=1}^3 \alpha_nA_n=0 \quad \sum_{n=1}^3 \alpha_nA_n^2= 1
\end{eqnarray}
So that finally
\begin{equation}
f^*(x\ge 2) = 2x \sum_{n=1}^3 \alpha_n(x-A_n)\ln \left (1-\frac{A_n}{x} \right ) 
\end{equation}
\begin{eqnarray}
f^*( x\le 2) = 2x\int_2^\infty d\zeta \int_\zeta ^\infty \frac{d\zeta '}{\kappa_2(\zeta ')}+\cr
2x \int_x^2 d\zeta \left \{ \int_2^\infty \frac{d\zeta}{\kappa_2(\zeta)} +\int_\zeta ^2 \frac{d\zeta '}{K_1(\zeta')}  \right \} 
\end{eqnarray}
\begin{equation}
f^*(x\le 2) =  2x \sum_{n=1}^3 \alpha_n(x-A_n)\ln(1-\frac{A_n}{2})+\frac{(x-2)^2}{4(\delta+1)}\
\end{equation}


%
%
 \bibliographystyle{ieeetr}
 \bibliography{biblio}
%

\end{document}